\documentclass[paper]{ieice}
\usepackage{amsthm}
\usepackage[dvipdfm]{graphicx}
\usepackage[fleqn]{amsmath}
\usepackage{newtxtext}
\usepackage[varg]{newtxmath}

\field{}
\title{Biometric Identification Systems With Both Chosen and Generated Secret Keys by Allowing Correlation}
\titlenote{This paper was presented in part at the 2020 International Symposium on Information Theory and Its Applications.
This study was supported in part by JSPS KAKENHI Grant Numbers JP20K04462, JP19J13686, JP18H01438, and JP17K00020.}
\authorlist{%
 \authorentry{Vamoua Yachongka}{m}{labelA}\MembershipNumber{}
 \authorentry{Hideki Yagi}{m}{labelB}\MembershipNumber{}
}
\affiliate[labelA]{The author is with Advanced Wireless \& Communication Research Center (AWCC),
The University of Electro-Communications,1-5-1 Chofugaoka, Tokyo 182-8585, Japan. Email: va.yachongka@uec.ac.jp}
\affiliate[labelB]{The author is with the Department of Computer and Network Engineering,
the University of Electro-Communications, Tokyo, 182-8585 Japan. Email: h.yagi@uec.ac.jp
}

\usepackage{cite}
\usepackage{amsmath,amssymb,amsfonts}
\usepackage{algorithmic}
\usepackage{graphicx}
\usepackage{textcomp}
\usepackage{xcolor}
\def\BibTeX{{\rm B\kern-.05em{\sc i\kern-.025em b}\kern-.08em
    T\kern-.1667em\lower.7ex\hbox{E}\kern-.125emX}}
\usepackage{url}
\usepackage{ifthen}
\usepackage{multicol}
\usepackage{latexsym}
\usepackage{wrapfig}
\usepackage{blindtext}
\usepackage{bm}
\usepackage{lipsum}
\usepackage{cuted}
\usepackage{amsthm}
\usepackage{ascmac}
\usepackage{lipsum}
\usepackage{mathtools}
\usepackage{cuted}
\usepackage{enumitem}
\usepackage{stfloats}
\usepackage{blindtext}
\usepackage{pifont}
\newtheorem{theorem}{Theorem}
\newtheorem{definition}{Definition}
\newtheorem{remark}{Remark}
\newtheorem{lemma}{Lemma}

\newtheorem{corollary}{Corollary}
\def\Var{{\textrm{Var}}\,}

\begin{document}
\maketitle
\begin{summary}
We propose a biometric identification system where the chosen- and generated-secret keys are used simultaneously, and investigate its fundamental limits from information theoretic perspectives. The system consists of two phases: enrollment and identification phases. In the enrollment phase, for each user, the encoder uses a secret key, which is chosen independently, and the biometric identifier to generate another secret key and a helper {data}. In the identification phase, observing the biometric sequence of the identified user, the decoder estimates index, chosen- and generated-secret keys of
the identified user based on the helper {data} stored in the system database. In this study, the capacity region of such system is characterized. In the problem setting, we allow chosen- and generated-secret keys to be correlated. As a result, by permitting the correlation of the two secret keys, the sum rate of the identification, chosen- and {generated-secret key} rates can achieve a larger value compared to the case where the keys {do not correlate}. Moreover, the minimum amount of the storage rate changes in accordance with both the identification and chosen-secret key rates, but that of the privacy-leakage rate depends only on the identification rate.
\end{summary}

\section{Introduction}
Biometrics based identification and authentication has been drawing public attention increasingly. It is renowned for the features of providing high security and convenience since biological data (bio-data) of our humankind such as finger, eyes, {and} palm, which cannot be forgotten or lost like password or smart-card, are used \cite{jainetal}. {\em Biometric identification systems} (BISs)
were first analyzed in \cite{OS} and \cite{willems}, and the identification capacity of the BIS was characterized in \cite{willems}.
In \cite{tuncel}, a constraint (lossless {coding}) on the helper data stored in a public database was added, and extended work of \cite{tuncel} can be found in \cite{tuncel2} to recover noisy reconstruction ({lossy coding}). Error exponents of the BIS {are} examined in \cite{vyisita2016} {based on} Arimoto's arguments
\cite{Arimoto1974}, and {in} \cite{zhou2018} from information-spectrum perspectives \cite{han2003}.

The BIS with estimating both user's index and secret key was first investigated in \cite{itw3}. In this work, two common BIS models, {\em generated secret} BIS (GS-BIS) model and {\em chosen secret} BIS (CS-BIS) model{,} were analyzed. In the GS-BIS model, a secret key is extracted from {a} bio-data sequence, while in the CS-BIS model, the secret key is chosen independently of it. Furthermore, {the GS-BIS model {in} the presence of an adversary was analyzed} \cite{kc}. Recently, Yachongka and Yagi characterized the capacity regions of more general models, where the noise in the enrollment phase was taken into account, in \cite{vy3} for the GS-BIS model via two auxiliary random variables (RVs) and in \cite{vy4} for both models via one auxiliary RV.
Another trend of studies is the BIS with one user, e.g., \cite{itw1}--\hspace{-.01em}\cite{onur}. {This system can be viewed as the source model with one-way communication only for the key-agreement problem considered in \cite{ahcr}, where eavesdropper has no side information related to the source sequence}. However, a privacy constraint, which was not imposed in \cite{ahcr}, was added in these studies.

In the previous studies such as \cite{itw3}, \cite{vy3}--\hspace{-.01em}\cite{onur}, the chosen- and generated-secret keys are assumed in the separate models, namely, {GS- and CS-BIS} models, respectively. Then, a rising question is when the two keys are used in the same system, how the chosen- and generated-secret key rates affect the fundamental performance of the BIS. One more thing is if a larger amount of information can be conveyed to the decoder {by allowing} these secret keys to be correlated. The answers to these questions have not yet been known, and {they are} not trivial from the results of the previous studies.

In this paper, the BIS model {in} a novel setting, where the chosen- and {generated-secret} keys are used together, is proposed, and we are {first characterize} the optimal trade-off of identification,
chosen- and {generated-secret key} rates under privacy and storage constraints for the discrete alphabet {sources}. Also, we allow the chosen- and {generated-secret} keys to be correlated at a certain level.
In the derivation, it seems hard to bound the privacy-leakage rate directly in the converse part, and we newly establish a lemma to overcome such difficulty. In addition, in {the} direct part, the degree of correlation for the two keys (proof by cases) {is carefully analyzed}. As a result, the characterization shows that identification, chosen- and {generated-secret key} rates are in a trade-off relation, and a larger sum of these rates is achievable. The minimum storage rate (storage space) requires to be larger {as} the identification and 
{chosen-secret key} rates increase, but it is not affected by the {generated-secret key} rate. Unlike the {storage} rate, the privacy-leakage rate {varies} in accordance with only the changes of the identification rate. As special cases, this result reduces to several known characterizations provided in previous {studies} \cite{vy4}, \cite{onur}, \cite{vy6}. Moreover, {extending the characterization for the system with discrete alphabets}, the optimal trade-off regions for binary and Gaussian sources are derived. {To illustrate the behaviors of the capacity region in the case where the two keys are allowed and {disallowed} to be correlated, some numerical calculations for the binary sources are given}.

The organization of this {paper} is as {follows}: We describe the basic settings of
{the} system model in {Sect.}\ \ref{sec2}. Our main result is presented in
{Sect.}\ \ref{sec3}, and the proof of {the} main result is given in the Appendixes.
Connections to the results in previous studies and examples are given in {Sect.}\ \ref{sec4}.
Finally, a short {concluding discussion} is given in {Sect.}\ \ref{sec5}.

\section{Notation and System Model} \label{sec2}

Calligraphic letter $\mathcal{A}$ stands for a finite {set}\footnote{{This assumption of finite alphabet is relaxed in Sect.\ 4.3 to consider continuous sources.}} and its cardinality is written as $|\mathcal{A}|$. Upper-case such {as} $A$ denotes a random variable (RV) taking values in $\mathcal{A}$ and lower-case $a \in \mathcal{A}$ denotes its realization. $A^n = (A_{1},\cdots ,A_{n})$ represents a string of RVs, taking values in $\mathcal{A}^n$, and subscripts represent the position of {an} RV in the string. $P_A(a)=\Pr[A = a]$, $a \in \mathcal{A}$, represents the probability distribution on $\mathcal{A}$. For integers $k$ and $t$ such that $k < t$, $[k:t]$ denotes the set $\{k,k+1,\cdots,t\}$. A partial sequence from the $k$th symbol to the $t$th symbol is represented by $c^t_k$. $\mathcal{T}^{(n)}_{\epsilon}(\cdot)$ denotes the strongly $\epsilon$-typical set \cite{cover}, \cite{GK}. $H_b(\cdot)$ is the binary entropy, and $H^{-1}_b(\cdot)$ is its inverse function. The $*$-operator is defined as $a*b=a(1-b)+(1-a)b$. $\ln x$ is the natural logarithm of $x>0$. $h(\cdot)$ is differential entropy.

\begin{figure}[!t]
 \centering
  \includegraphics[width =85mm]{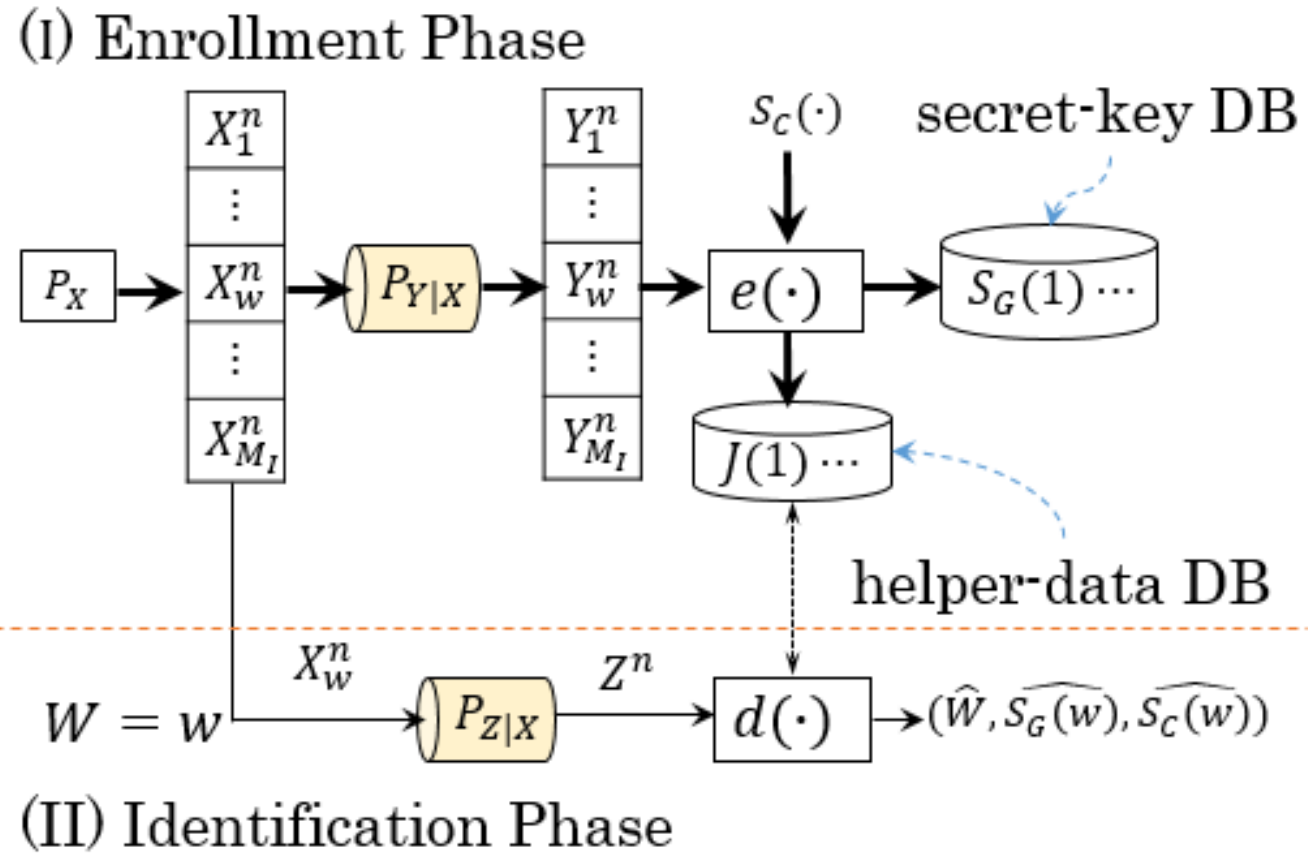}
   \vspace{-3mm}
 \caption{System model}
 \vspace{-4mm}
 \label{fig22}
\end{figure}

The system model considered in this paper {is} illustrated in Fig.\ \ref{fig22}.
$P_X$, $P_{Y|X}$, and $P_{Z|X}$ denote {the} biometric source, enrollment channel, and identification channel, respectively.
$P_{Y|X}$ and $P_{Z|X}$ are discrete memoryless channels (DMCs).
Let $\mathcal{I} = [1:M_I]$ and $\mathcal{J}= [1:M_J]$ be the sets of user's {indexes} and helper {data}.
{For each user $(i\in\mathcal{I})$, the chosen- and generated-secret keys of $i$ are denoted by $S_C(i)$ and $S_G(i)$, respectively}. Let $\mathcal{S}_C = [1:M_C]$ and $\mathcal{S}_G= [1:M_G]$ be the sets of the chosen- and generated-secret keys. Lowercase letters $s_C(i) \in \mathcal{S}_C$ and $s_G(i) \in \mathcal{S}_G$ stand for their realizations.

The BIS consists of two phases: {\em Enrollment Phase} and {\em Identification Phase}. {Random vector $X^n_i$ denotes the source sequences of user $i \in \mathcal{I}$ and each symbol of $X^n_i$ is generated independently and identically distributed (i.i.d.)\ from $P_X$. Random vectors $Y^n_i$ and $Z^n$ are the outputs of the enrollment {channel $P_{Y|X}$} and {the} identification channel {$P_{Z|X}$}, respectively, with having $X^n_i$ as input}. Assume that the chosen-secret key is uniformly distributed on {$\mathcal{S}_C$}, i.e.,
$
P_{S_C(i)}({s}) = \frac{1}{M_C}
$
for all $i \in \mathcal{I}$ and {${s} \in \mathcal{S}_C$}.
In the Enrollment Phase, observing $Y^n_i$ and $S_C(i)$,
the encoder {$e$} generates
the pair
$
(J(i),S_G(i)) = {e}(Y^n_i,S_C(i)),
$
where $J(i)$ is called {a} {\em helper {data}} and {takes} values {in} $\mathcal{J}$.
This operation {is repeated} for all users. The $J(i)$ and $S_G(i)$ are stored at position $i$ in the helper-{data} and secret-key databases (DBs), respectively. The key DB is {assumed} to be installed in a secure place. We denote the helper-{data} DB $\{J(1),{\cdots},J(M_I)\}$ as $\bm{J}$ {for brevity}. Let $W$ and $\widehat{W}$ denote the {indexes} of the identified user and its estimate, respectively.
In the Identification Phase,
seeing $Z^n$, the decoder reconstructs index and secret keys by
$
(\widehat{W},\widehat{S_C(w)},\widehat{S_G(w)}) = {d(Z^n,\bm{J})}.
$

Note that the task of the decoder is to estimate the index and the secret keys assigned to each user.
One possible application of this model is the BIS that supports two-factor authentication if one
thinks of using the estimated index {to claim} who the identified user is, and the two estimated secret keys for
the first and second rounds of authentications. However, the use of the estimated values depends on user's purpose, and such {an}
argument is not provided in this paper as it goes beyond the scope of our focus.

\vspace{-1mm}
\section{Problem Formulation and Main Results} \label{sec3}
Let {$E(W)$ and $\widehat{E(W)}$} represent the tuples {$(W,S_C(W),S_G(W))$ and $(\widehat{W},\widehat{S_C(W)},\widehat{S_G(W)})$}, respectively.

\begin{definition} \label{def11}
A tuple of identification, chosen- and generated-{secret key}, storage, and privacy-leakage rates
$(R_I,R_C,R_G,R_J,R_L) \in \mathbb{R}^5_+$ is
said to be $\Gamma$-achievable {for $\Gamma > 0$} if for any
$\delta > 0$ and large enough $n$ there exist pairs of encoders and decoders that satisfy
\vspace{-2mm}
\begin{align}
   \textstyle\max_{i \in \mathcal{I}}\Pr\{\widehat{E(W)}\neq E(W) |W=i\} &\leq  \delta, \label{4a} \\
   \textstyle\frac{1}{n}\log M_I &\geq R_I - \delta{,} \label{4b} \\
   \textstyle\frac{1}{n}\log M_C &\geq R_C - \delta{,} \label{4c} \\
   \textstyle\min_{i \in \mathcal{I}}\frac{1}{n}H(S_G(i)) &\geq R_G - \delta, \label{4e} \\
  \textstyle \frac{1}{n}\log{M_J} &\leq R_J + \delta, \label{4d} \\
   \textstyle\max_{i \in \mathcal{I}}\frac{1}{n}I(X^n_i;J(i)) &\leq R_L + \delta, \label{4f} \\
   \textstyle\max_{i \in \mathcal{I}}\frac{1}{n}I(S_C(i);S_G(i)) &\leq \Gamma, \label{4g} \\
   \textstyle\max_{i \in \mathcal{I}}\frac{1}{n}I(S_C(i),S_G(i);J(i)) &\leq \delta. \label{4h}
\end{align}
Moreover, define $\mathcal{R}(\Gamma)$ as {the closure of the set of all $\Gamma$-achievable rate tuples for the BIS},
called the $\Gamma$-capacity region.
\end{definition}

{One of the main results in} this paper is presented below.
\begin{theorem} \label{th1}
The $\Gamma$-capacity region of the BIS is given by
\vspace{-2mm}
\begin{align}
\hspace{-7mm}\mathcal{R}(\Gamma) = \Big\{&(R_I,R_C,R_G,R_J,R_L)\in \mathbb{R}^5_+:~R_I + R_C \leq I(Z;U), \nonumber \\
&R_I + R_G \le I(Z;U), \nonumber \\
&R_I + R_C + R_G \leq I(Z;U) + \min\{\Gamma,R_C,R_G\}, \nonumber \\
&R_J \geq I(Y;U) - I(Z;U) + R_I + R_C, \nonumber \\
&R_L \geq I(X;U) - I(Z;U) + R_I~~~~\mathrm{for~some}~U~\mathrm{s.t.} \nonumber \\
&Z-X-Y-U~{\rm with}~|\mathcal{U}| \leq |\mathcal{Y}| + 2\Big\}. \label{theorem1}
\end{align}
\vspace{-2mm}
\qed
\end{theorem}
\begin{remark}
For the convenience of analysis, in this paper, we focus only on the case where
the condition of {(normalized)} secrecy-leakage (cf.\ \eqref{4h}) is
imposed under a week secrecy criterion. The achievability proof of Theorem \ref{th1} makes use of random coding arguments \cite{cover}.
However, a privacy amplification technique developed in \cite{wataoha2010} based on information-spectrum approaches \cite{han2003}
can be used to show that the secrecy-leakage under a strong secrecy criterion is achievable. More specifically,
combining \cite[Lemma 12]{wataoha2010} with \cite[Lemma 3]{Naito2008},
the {(unnormalized)} secrecy-leakage of the BIS can be bounded by a negligible amount
regardless of the bloch length $n$, and the capacity regions of both cases
coincide. The readers may refer to \cite[Appendix A]{vyo2022} for detailed analysis.
\end{remark}
Using a similar technique shown in
{\cite[Sect.\ IV-A]{itw3}}, one can easily check that $\mathcal{R}(\Gamma)$ is a convex region.
For the detailed proof of Theorem \ref{th1}, see Appendix A.

\begin{figure}[!ht]
\vspace{-4mm}
    \centering
    \includegraphics[scale=0.4]{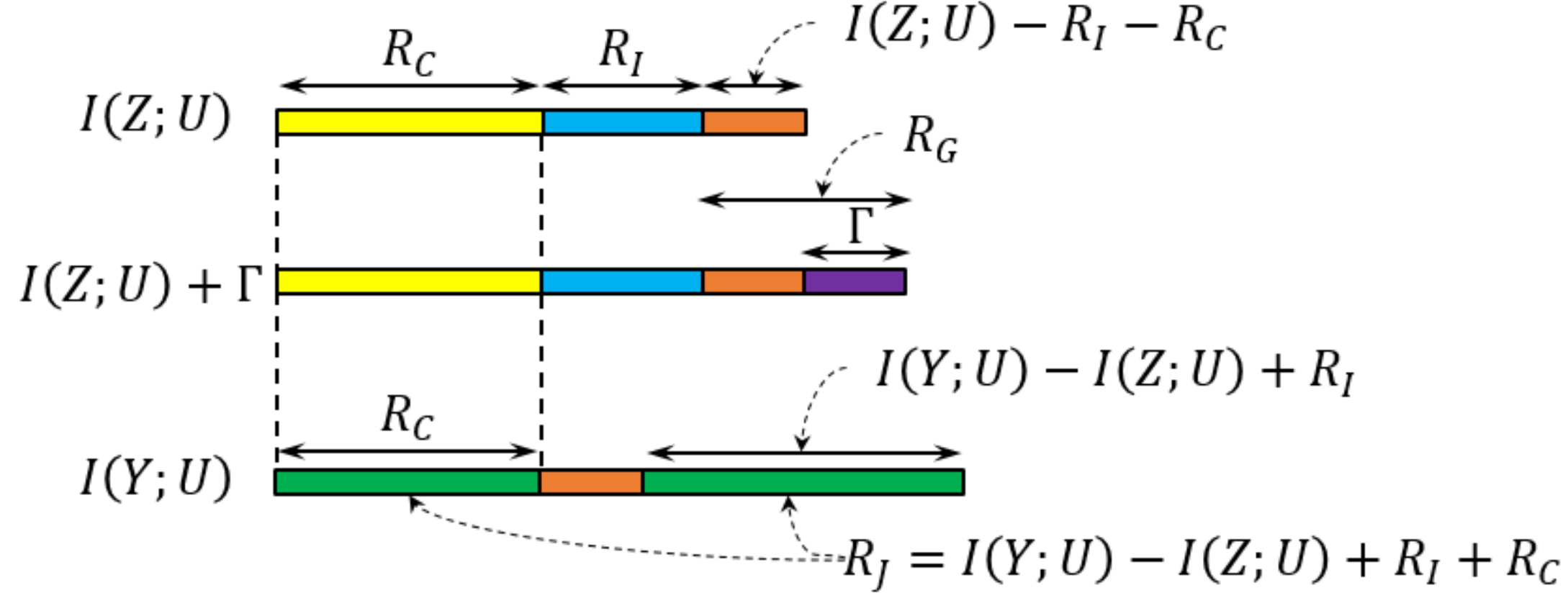}
    \caption{An explanation of {the optimal values of each rate} in Theorem \ref{th1} for the case where
     $\Gamma \le {\min\{R_C,R_G\}}$.}
    \label{Thm4}
\vspace{-4mm}
\end{figure}

In our setting, the decoder is required to reconstruct
the user index and both secret keys. In \cite{vy6}, the authors showed that the sum of identification, generated- and chosen-{secret key} rates cannot
be larger than $I(Z;U)$ if the right-hand side of \eqref{4g} is replaced with a negligible amount
(vanishing secrecy-leakage rate). However, since we permit the chosen- and generated-secret keys to be correlated {(non-vanishing secrecy-leakage rate)},
the maximum recognizable value for the sum of these rates is $I(Z;U) + \Gamma$, which can exceed the result in \cite{vy6}.
(cf.\ the middle band graph in Fig.\ \ref{Thm4})
More precisely, $R_I$ and $R_C$ can be any value in the range of $[0,I(Z;U)]$ under a constraint that their sum should be less than $I(Z;U)$ as shown in the top band graph. The {optimal value} of the generated-secret key rate $R_G$ is $I(Z;U) + \Gamma - (R_I + R_C)$, which could be originally achieved at most $I(Z;U) - (R_I + R_C)$
for the case of vanishing secrecy-leakage rate \cite{vy6}.

The {optimal} amount of the storage rate is shown in the bottom band graph of Fig.\ \ref{Thm4}, which is the sum of the two green parts
($I(Y;U) - I(Z;U) + R_I + R_C$). The storage rate $R_J$ depends on both the identification rate $R_I$ and the chosen-{secret key} rate $R_C$,
and this value is larger than the one derived in \cite[Theorem 1]{vy4}.
This is because the information related to the chosen-secret key is needed {to} be stored in the DB, so that it can be reconstructed at the decoder reliably.

Since the minimum value of the storage rate of Theorem \ref{th1} is larger than
the one in \cite[Theorem 1]{vy4}, it is {somehow} expected that a larger storage rate might lead to leaking a more amount of user's privacy. {Nevertheless}, the minimum amount of the privacy-leakage rates of Theorem \ref{th1} and \cite[Theorem 1]{vy4} is bounded by the same quantity, which is $I(X;U)-I(Z;U)+R_I$. Obviously, the chosen-secret key rate is not {involved}, and only the changes of the identification rate affect the minimum required amount of the privacy-leakage rate. The reason is because the bits related to the chosen-secret key stored in the DB is made perfectly confidential, e.g., using the one-time pad operation, and this portion makes no contribution to the privacy-leakage.

\section{Special Cases and Examples} \label{sec4}
\subsection{Connections to Previous Results}
We can check that Theorem \ref{th1} covers the several results provided in previous studies.
For instance, in the case where the chosen- and generated-secret key do not correlated ($\Gamma = 0$), $\mathcal{R}(\Gamma)$ naturally reduces to the one given in \cite[Theorem 1]{vy6}. When there are no provision of secret {keys} ($R_C = 0$)
and no allowance of secret-key correlation ($\Gamma = 0$),
$\mathcal{R}(\Gamma)$ coincides with the one given in \cite[Theorem 1]{vy4}. In the case of no {secret key} generation
($R_G = 0$) and no allowance of secret-key correlation ($\Gamma = 0$), the capacity region,
denoted by $\mathcal{R}'$ in this case, is given in the following corollary.
\begin{corollary}
\begin{align}
\hspace{-3mm} \mathcal{R}' = \Big\{(&R_I,R_C,R_J,R_L)\in \mathbb{R}^4_+:~~R_I + R_C \leq I(Z;U),\nonumber \\
&R_J \geq I(Y;U) - I(Z;U) + R_I + R_C,\nonumber \\
&R_L \geq I(X;U) - I(Z;U) + R_I~~~~\mathrm{for~some}~U~\mathrm{s.t.} \nonumber \\
&Z-X-Y-U~{\rm with}~|\mathcal{U}| \leq |\mathcal{Y}| + 2\Big\}. \label{cor1}
\end{align}
\qed
\end{corollary}
Although the expression of $\mathcal{R}'$ and the one given in \cite[Theorem {2}]{vy4} are different, it can be checked that both are identical. One can easily see that $\mathcal{R}'$ is contained in the region of
\cite[Theorem 2]{vy4} due to the range of $R_J$. For proving the opposite relation, we choose a new test channel
$P_{U'|U}$ satisfying that $R_I+R_C = I(U';Z)$. We can pick such channel since $I(Z;U) \ge I(Z;U') \ge 0$ and $I(Z;U')$
is a continuous function {of $P_{U'|U}$}. The bounds {on} the storage and privacy-leakage rates become
\begin{align}
    R_J &\ge I(Y;U) - I(Z;U) + I(Z;U') \nonumber \\
    &\overset{\mathrm{(a)}}{\ge} I(Y;U') - I(Z;U') + I(Z;U')= I(Y;U') \label{rjr} \\
    R_L &\ge I(X;U) - I(Z;U) + R_I \nonumber \\
    & \overset{\mathrm{(b)}}{\ge} I(X;U') - I(Z;U') + R_I, \label{rlr}
\end{align}
where (a) and (b) follow from the face that $I(Y;U|Z) \ge I(Y;U'|Z)$ and $I(X;U|Z) \ge I(X;U'|Z)$, respectively.
Hence, there always exists an auxiliary $U'$ where an achievable rate tuple $(R_I,R_C,R_J,R_L)$ in the region of
\cite[Theorem 2]{vy4} is also included in $\mathcal{R}'$.

Moreover, in the case where we set $R_I$ and $\Gamma$ to be zero (single user case without allowing
secret-key correlation), the capacity region, denoted by $\mathcal{R}''$
in this case, is obtained.
\begin{corollary}
\begin{align}
\hspace{-2mm}\mathcal{R}'' = \Big\{(&R_C,R_G,R_J,R_L)\in \mathbb{R}^4_+: R_C + R_G \leq I(Z;U),\nonumber \\
&R_J \geq I(Y;U) - I(Z;U) + R_C,\nonumber \\
&R_L \geq I(X;U) - I(Z;U)~~~~\mathrm{for~some}~U~\mathrm{s.t.} \nonumber \\
&Z-X-Y-U~{\rm with}~|\mathcal{U}| \leq |\mathcal{Y}| + 2\Big\}. \label{cor2}
\end{align}
\qed
\end{corollary}
When $R_C = 0$ (no provision of secret {keys}), one can easily see that $\mathcal{R}''$ is equivalent to
the one given in \cite[Theorem 1]{onur}. Moreover, in {the} case $R_G = 0$ (no generation of secret {keys}),
it can also be shown that $\mathcal{R}''$ matches the region provided in \cite[Theorem 2]{onur}
by a similar argument {for} proving that $\mathcal{R}'$ and the region of \cite[Theorem 2]{vy4} are the
same (cf.\ \eqref{rjr} and \eqref{rlr}).

\subsection{Examples of Binary Sources}

In this section, a numerical example of the rate region of the BIS for {a} binary hidden source is given. We consider the case where $P_X(0)=P_{X}(1)=0.5$, {and the enrollment channel $P_{Y|X}$ and the identification channel $P_{Z|X}$ of the systems are binary symmetric channels with crossover probabilities $0 \le p_E \le 0.5$ and $0 \le p_D \le 0.5$, respectively.} First, we simplify the capacity region for this case by applying Mrs.\ {Gerber's} Lemma (MGL) \cite{wyner1973} twice into two opposite directions. The simplification of the rate region of the BIS with one user for binary hidden sources was given in \cite{onur}, too. However, by introducing an additional parameter, our deriving method is {simpler than} the one shown in \cite{onur}. In the right-hand side of \eqref{theorem1}, we have that
\begin{align}
    I(Z;U) &= 1 - H(Z|U), \nonumber \\
    I(Y;U) - I(Z;U) &= H(Z|U) - H(Y|U),  \nonumber \\
    I(X;U) - I(Z;U) &= H(Z|U) - H(X|U). \label{hzuxuri}
\end{align}

From {the} above relations, it indicates that to simplify the capacity region, we must maximize $H(Y|U)$
and minimize $H(Z|U)$ for fixed $H(X|U)$. First, observe that since $1\ge H(X|U) \ge H(X|Y) = H_b(p_E)$,
there must exist {a} $\gamma$ satisfying that
\begin{align}
H(X|U) = H_b(\gamma*p_E),
\end{align}
where $\gamma \in [0,0.5]$.
By applying MGL to the Markov chain $U-X-Z$, we have
\begin{align}
    \hspace{-5mm}H(Z|U) \ge H_b(H^{-1}_b(H(X|U))*p_D) = H_b(\gamma*p_E*p_D). \label{hzuped}
\end{align}
Again, in {the} opposite direction, if MGL is applied to the Markov chain $U-Y-X$, it follows that
\begin{align}
    H(X|U) \ge H_b(H^{-1}_b(H(Y|U))*p_E). \label{hxupe}
\end{align}
As $H(X|U) = H_b(\gamma*p_E)$, \eqref{hxupe} yields that
\begin{align}
    H_b(\gamma*p_E) \ge H_b(H^{-1}_b(H(Y|U))*p_E), 
\end{align}
and thus
\begin{align}
    \gamma*p_E \ge H^{-1}_b(H(Y|U))*p_E.
\end{align}
Therefore, we obtain
\begin{align}
    H(Y|U) \le H_b(\gamma). \label{hyualpha}
\end{align}
In \eqref{hzuped} and \eqref{hyualpha} for binary symmetric {$P_{U|Y}$} with crossover probability $\gamma$,
the minimum $H(Z|U)=H_b(\gamma*p_E*p_D)$ and the maximum $H(Y|U)=H_b(\gamma)$ are achieved.
Therefore, the following {theorem} is obtained. We denote the $\Gamma$-capacity region for {binary} sources as
$\mathcal{R}_b(\Gamma)$.

\begin{theorem}
For binary hidden sources, \eqref{theorem1} reduces to
\vspace{-2mm}
\begin{align}
\hspace{-5mm}\mathcal{R}&_b(\Gamma) = \bigcup_{\gamma \in [0,0.5]}\Big\{(R_I,R_C,R_G,R_J,R_L)\in \mathbb{R}^5_+: \nonumber \\
    &R_I + R_C \le 1 - H_b(\gamma*p_E*p_D), \nonumber \\
    &R_I + R_G \le 1 - H_b(\gamma*p_E*p_D), \nonumber \\
    &R_I + R_C + R_G \le 1-H_b(\gamma*p_E*p_D) \nonumber \\
    &~~~~~~~~~~~~~~~~~~~~~~~~~~~~+ {\min\{\Gamma,R_C,R_G\}}, \nonumber \\
    &R_J \ge H_b(\gamma*p_E*p_D) - H_b(\gamma) + R_I + R_C, \nonumber \\
    &R_L \ge {H_b(\gamma*p_E*p_D) - H_b(\gamma*p_E) + R_I}~\Big\}. \label{cor3}
\end{align}
\qed
\end{theorem}

{We calculate the rate region above under two settings; both $R_I$ and $\Gamma$ are zero, and $R_I$ is zero, but $\Gamma$ is a positive value. The crossover probabilities of enrollment and identification channels are set to be $p_E=0.03$ and $p_D=0.1$, respectively, which are close to the actual transition probabilities {of} the channels in the real-life systems [10], [17]. To recap how secret-key correlation affects the rate region of BIS, we further fix the chosen-secret key rate to be $I(Z;U)$ and $\frac{1}{2}I(Z;U)$, and investigate the optimal values of the generated-secret key and storage rates under {these} two settings. The numerical results are shown in {Figures 3(a) and 3(b)}. In both figures, the graphs with blue circles and red asterisks represent the boundaries of the pair $(R_J,R_G)$ in the case where $\Gamma = 0$ and $\Gamma = 0.2$, respectively.}

{{In Figure 3(a)}, one can see that when $\Gamma = 0$, no positive generated-secret key rate is achievable. On the other hand, when $\Gamma = 0.2$, it is possible to generate secret keys at a positive rate. This is implementable by sharing information bits of the chosen-secret keys to make the generated-secret keys. However, {the sharing} quantity {cannot} exceed 0.2, the maximum {value} that the two keys are allowed to be correlated. Similarly, {in Figure 3(b)}, the calculated result shows that when $\Gamma = 0.2$, the BIS provides {a} higher generated-secret key rate.}

{From these behaviours, it is obvious that by allowing correlation, the greater sum rates of chosen- and generated-secret keys are attainable. In other words, the capacity region of the BIS with permitting the secret-key correlation is larger {than the one} with {the} {disallowed} correlation case.}

\begin{figure*}[!t]
\setcounter{figure}{2} 
\centering
\begin{minipage}{.5\textwidth}
  \centering
  \includegraphics[width=.95\linewidth]{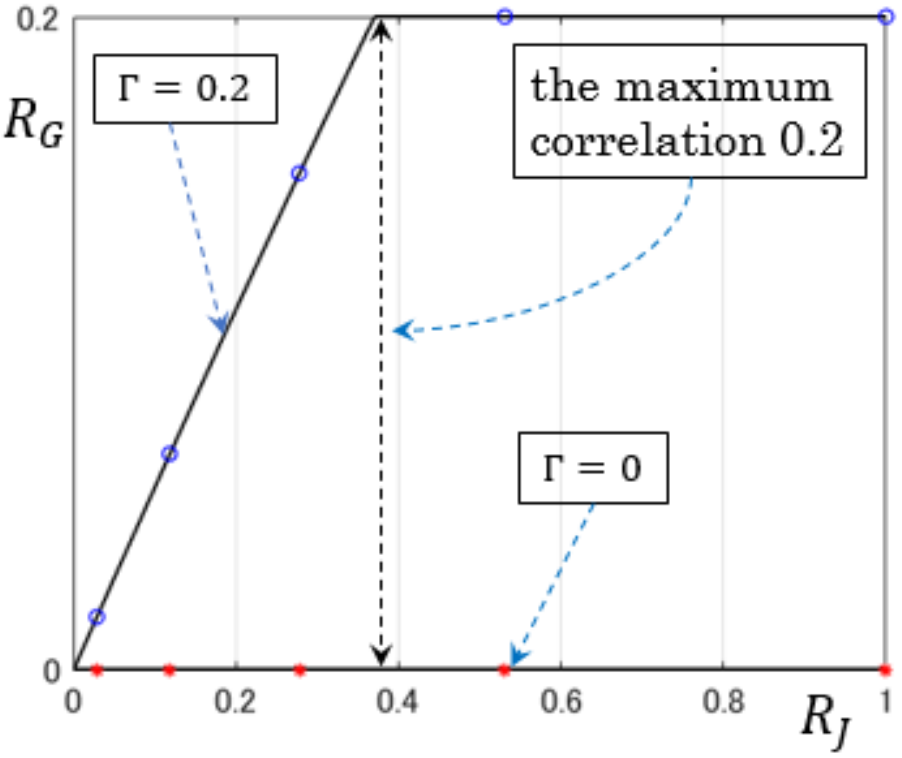}
  \\
  {(a)~~~{$R_C = I(Z;U)$}}
\end{minipage}%
\begin{minipage}{.5\textwidth}
  \centering
  \includegraphics[width=.95\linewidth]{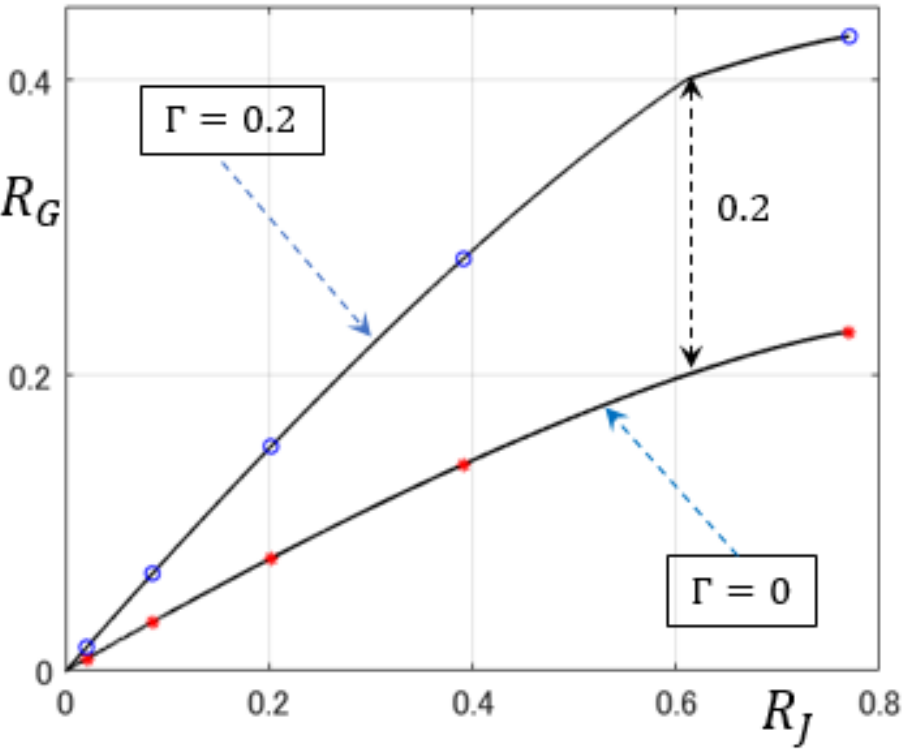}
  \\
  {(b)~~~{$R_C=\frac{1}{2}I(Z;U)$}.}
\end{minipage}
\vspace{2mm}
\caption{{{The left-hand side (a) and right-hand side (b) figures depict the boundaries of the rate pair $(R_J,R_G)$ for $R_C = I(Z;U)$ and $R_C=\frac{1}{2}I(Z;U)$, respectively, under settings where $\Gamma =0$ and $\Gamma = 0.2$.}}}
\label{rlrc}
\end{figure*}

\subsection{{Examples of Gaussian Sources}} \label{egs}
We discuss the case of Gaussian RVs $(X,Y,Z)$, where the alphabets are continuous. Assume that biometric sources $X \sim \mathcal{N}(0,1)$, where $\mathcal{N}(0,1)$ is {the standard} Gaussian {distribution} with mean zero and variance one. The enrollment and identification channels are modeled as
$Y = \rho_1X + N_1,$ and $Z = \rho_2X + N_2,$ respectively, where $|\rho_1|,|\rho_2| < 1$, and $N_1 \sim \mathcal{N}(0,1-\rho^2_1)$ and $N_2 \sim \mathcal{N}(0,1-\rho^2_2)$ are independent of each other and other RVs.
{By a similar procedure of deriving Theorem \ref{th1}, it can be proved that the single-letter expression in \eqref{theorem1} also holds for Gaussian sources and channels. However, since the alphabets, e.g, $\mathcal{X},\mathcal{Y},\mathcal{Z}$, and $\mathcal{U}$, are unbounded, the region is not directly computable. In this section, we aim to derive a parametric form for {the} Gaussian case via the constraints in \eqref{theorem1}}.

Let $\mathcal{R}_g(\Gamma)$ denote
the $\Gamma$-capacity region for Gaussian sources {and channels}.
\begin{theorem} \label{guassian}
For i.i.d. Gaussian sources, the {$\Gamma$-capacity region} of the BIS is given by
\begin{align}
\hspace{-5mm}\mathcal{R}_g&(\Gamma) = \bigcup_{\alpha \in (0,1]}\Big\{(R_I,R_C,R_G,R_J,R_L)\in \mathbb{R}^5_+: \nonumber \\
&R_I + R_C \leq \frac{1}{2}\ln\left(\frac{1}{\alpha\rho^2_1\rho^2_2 + 1 - \rho^2_1\rho^2_2}\right), \nonumber \\
&R_I + R_G \leq \frac{1}{2}\ln\left(\frac{1}{\alpha\rho^2_1\rho^2_2 + 1 - \rho^2_1\rho^2_2}\right), \nonumber \\
&R_I + R_C + R_G \leq \frac{1}{2}\ln\left(\frac{1}{\alpha\rho^2_1\rho^2_2 + 1 - \rho^2_1\rho^2_2}\right) \nonumber \\
&~~~~~~~~~~~~~~~~~~~~~~~~~~~~~~~~+ {\min\{\Gamma,R_C,R_G\}}, \nonumber \\
&R_J \geq \frac{1}{2}\ln\left(\frac{\alpha\rho^2_1\rho^2_2 + 1 - \rho^2_1\rho^2_2}{\alpha}\right) + {R_I + R_C}, \nonumber \\
&R_L \geq \frac{1}{2}\ln\left(\frac{\alpha\rho^2_1\rho^2_2 + 1 - \rho^2_1\rho^2_2}{\alpha\rho^2_1 + 1 - \rho^2_1}\right)
+ R_I~~\Big\}. \label{theoremg}
\end{align}
\qed
\end{theorem}
For proving \eqref{theoremg}, a technique of the converted system introduced in \cite{YYO2021a} plays an important role. The technique tell us that the joint density of the system ($Y = \rho_1X + N_1,~Z = \rho_2X + N_2$) is equivalent to that of
\begin{align}
    X &= \rho_1 Y + N'_1, \label{xyn} \\
    Z &= \rho_2 X + N_2 \label{zyn}
\end{align}
with {$N'_1 \sim \mathcal{N}(0,1-\rho^2_1)$}. To derive the parametric form, we will make use of the relations \eqref{xyn} and \eqref{zyn} to prove \eqref{theoremg} instead of the relation of the original system.

\medskip
\noindent{\em Proof of Achievability}:~~~Let $0< \alpha \le 1$. Also, let $U$ be an auxiliary Gaussian RV with mean zero and
variance $1-\alpha$, i.e., $U \sim \mathcal{N}(0,1-\alpha)$, and $\Phi \sim \mathcal{N}(0,\alpha)$. Choose $Y = U + \Phi$. Put this $Y$ into \eqref{xyn},
we obtain $X = \rho_1 U + \rho_1\Phi + N'_1$, and now put $X$ into \eqref{zyn}, we have that
$Z = \rho_1\rho_2 U + \rho_1\rho_2 \Phi + \rho_2N'_1 + N_2$. Observe that $\Var[\rho_1\Phi + N'_1] = \rho^2_1\Var[\Phi] + \Var[N'_1] = \alpha\rho^2_1 + 1-\rho^2_1$
and $\Var[\rho_1\rho_2 \Phi + \rho_2N'_1 + N_2] = \rho^2_1\rho^2_2\Var[\Phi] + \rho^2_2\Var[N'_1] + \Var[N_2] = \alpha\rho^2_1\rho^2_2 + 1-\rho^2_1\rho^2_2$.
Now it can be calculated that $I(Y;U) = \frac{1}{2}\ln\left(\frac{1}{\alpha}\right)$, $I(X;U) = \frac{1}{2}\ln\left(\frac{1}{\alpha\rho^2_1 + 1-\rho^2_1}\right)$,
and $I(Z;U) = \frac{1}{2}\ln\left(\frac{1}{\alpha\rho^2_1\rho^2_2 + 1-\rho^2_1\rho^2_2}\right)$. Furthermore, it is not difficult to see that
$I(Y;U)-I(Z;U) = \frac{1}{2}\ln\left(\frac{\alpha\rho^2_1\rho^2_2 + 1-\rho^2_1\rho^2_2}{\alpha}\right)$ and
$I(X;U)-I(Z;U) = \frac{1}{2}\ln\left(\frac{\alpha\rho^2_1\rho^2_2 + 1-\rho^2_1\rho^2_2}{\alpha\rho^2_1 + 1-\rho^2_1}\right)$.
From \eqref{theorem1}, it is easy to see that the right-hand side of \eqref{theoremg} is achievable for this choice of $U$.

\medskip
\noindent{\em Proof of Converse}:~~~Similar to the development of \eqref{hzuxuri}, for Gaussian sources, mutual information on the right-hand side of \eqref{theorem1} can be expanded as $I(Z;U) = \frac{1}{2}\log(2 \pi e) - {h(Z|U)}$, $I(Y;U)-I(Z;U) = {h(Z|U)-h(Y|U)}$, and $I(X;U)-I(Z;U) = {h(Z|U) - h(X|U)}$. {To obtain the outer bound of Gaussian sources for the region \eqref{theorem1}, we need to derive the optimal lower bound on $h(Z|U)$ and the optimal upper bound on $h(Y|U)$ under a fixed condition of $h(X|U)$}.

Now let us fix
\begin{align}
h(X|U) = \frac{1}{2}\ln\left(2 \pi e (\alpha\rho^2_1 + 1-\rho^2_1)\right) \label{hxu123}
\end{align}
for $0 < \alpha \le 1$. This is an appropriate setting since we have that $\frac{1}{2}\ln\left(2 \pi e \right) = h(X) \ge h(X|U) \ge h(X|Y) = \frac{1}{2}\ln\left(2 \pi e (1-\rho^2_1)\right)$.
By applying the conditional entropy power inequality (EPI) \cite{bergmans1974} to \eqref{zyn}, it follows that
\begin{align}
\hspace{-5mm}e^{2h(Z|U)} &\ge e^{2h(\rho_2X|U)} + e^{2h(N_2|U)} = \rho^2_2e^{2h(X|U)} + e^{2h(N_2)} \nonumber \\
&= \rho^2_2 (2 \pi e(\alpha\rho^2_1 + 1-\rho^2_1)) + 2 \pi e(1 - \rho^2_2) \nonumber \\
&= 2 \pi e(\alpha\rho^2_1\rho^2_2 + 1- \rho^2_1\rho^2_2),
\end{align}
and thus,
\begin{align}
h(Z|U) \ge \frac{1}{2}\ln\left(2 \pi e(\alpha\rho^2_1\rho^2_2 + 1- \rho^2_1\rho^2_2)\right). \label{hzu123}
\end{align}
On the other hand, again applying the conditional EPI to \eqref{xyn}, we have that
\begin{align}
\hspace{-5mm}e^{2h(X|U)} &\ge e^{2h(\rho_1Y|U)} + e^{2h(N'_1|U)} = \rho^2_1e^{2h(Y|U)} + e^{2h(N'_1)}. \label{xuyun111}
\end{align}
Next, {plugging} the value of $h(X|U)$ into \eqref{xuyun111}, it follows that
\begin{align}
2 \pi e (\alpha\rho^2_1 + 1-\rho^2_1) \ge \rho^2_1 e^{2h(Y|U)} + 2 \pi e(1 - \rho^2_1),
\end{align}
and finally, it can be concluded that
\begin{align}
e^{2h(Y|U)} \le 2 \pi e \alpha~~~{\rm or }~~~h(Y|U) \le \frac{1}{2}\ln\left(2 \pi e \alpha\right). \label{hyu123}
\end{align}
Now using \eqref{hxu123}, \eqref{hzu123}, and \eqref{hyu123}, one can see that the right-hand side of \eqref{theorem1} is contained in the right-hand side of \eqref{theoremg}, and thus the converse proof is completed.
\qed

\section{Conclusion and {Future Work}} \label{sec5}
In this study, we proposed the BIS with both chosen- and generated-secret keys, and characterized the capacity region among identification, chosen- and generated-secret key, storage, and privacy-leakage rates for the system. The {characterization shows} that identification, chosen- and generated-{secret key} rates are in a trade-off relation, and by permitting the correlation of the two secret keys, a larger sum of these rates is achievable. In addition, larger memory space for the DB is required when the sum of identification and chosen-{secret key} rates increases. Unlike the storage rate, only the identification rate contributes to the minimum required amount of the privacy-leakage rate, but the chosen-secret key rate does not. As special cases, this {characterization} reduces to the results seen in \cite{vy4} and \cite{onur}.


{For future work, an interesting problem is application of rate-distortion theory to the BIS. In \cite{tuncel2}, though lossy source coding was applied to the BIS, however, the model considered in the paper only dealt with user's identification and requirements on secrecy and privacy were not imposed. Therefore, there are still rooms for discussions about lossy source coding for the BIS in which secrecy and privacy constraints are taken into account. Another problem is to extend the result in Sect.\ 4.3 to vector Gaussian sources and channels, and clarify the capacity region of the BIS. Actually, as it was mentioned in Sect.\ 1, there are similarities between the BIS and the key-agreement model. To obtain the capacity region of the BIS for vector Gaussian sources, the technique used in [28] for analyzing the optimal trade-off of the key-agreement model may be useful}.


\section*{Appendix A: Proof of Theorem \ref{th1}}
In this appendix, we provide the detailed proof of Theorem \ref{th1}, the result for discrete sources and channels.
\subsection*{A. Converse Part}

We consider a more relaxed case where RV $W$ is uniformly distributed on $\mathcal{I}$,
and (\ref{4a}), \eqref{4e}, and \eqref{4f}--\eqref{4h} are replaced by
\begin{align}
   \textstyle\Pr\{\widehat{E(W)}\neq E(W)\} &\leq  \delta, \label{4ca} \\
   \textstyle \frac{1}{n}H(S_G(W)|W) &\geq R_G - \delta, \label{4ce} \\
   \textstyle \frac{1}{n}I(X^n_W;J(W)|W) &\leq R_L + \delta, \label{4cf} \\
   \textstyle \frac{1}{n}I(S_C(W);S_G(W)|W) &\leq \Gamma, \label{4cg} \\
   \textstyle \frac{1}{n}I(S_C(W),S_G(W);J(W)|W) &\leq \delta, \label{4ch}
\end{align}
respectively. Note that the capacity region of the BIS {under the average error criterion is fundamentally larger than the capacity region evaluated under the maximum error criterion.} We demonstrate that even this case, the outer bound of the capacity region coincides with its inner bound derived under the circumstance that the prior distribution of $W$ is unknown.

We assume that a rate tuple $(R_I,R_C,R_G,R_J,R_L)$ is achievable, implying that there exists a pair of an encoder and  a decoder satisfying \eqref{4b}, \eqref{4c}, \eqref{4d}, and \eqref{4ca}--\eqref{4ch}. First, we provide some useful lemmas. For $t \in [1:n]$, we define an auxiliary RV $U_t = (Z^{t-1},T(W))$, where
$
T(W) = (J(W),S_C(W),S_G(W),W).
$
We denote strings of RVs {by} $X^n_W  = (X_1(W) ,\cdots,X_n(W))$ and
$Y^n_W  = (Y_1(W) ,\cdots,Y_n(W))$. {Also, the partial random sequences $X^{t}(W)=(X_1(W),\cdots,X_{t}(W))$ and $Y^{t}(W)=(Y_1(W),\cdots,Y_{t}(W))$ represent strings of RVs from the first to $t$th positions in the sequences $X^n_W$ and $Y^n_W$ of user $W$, respectively.}
\begin{lemma} \label{lemma1}
The following Markov chains hold:
\begin{align}
Z^{t-1} - (Y^{t-1}(W),T(W)) - Y_t(W), \label{zytw} \\
Z^{t-1} - (X^{t-1}(W),T(W)) - X_t(W). \label{zxtw}
\end{align}
\end{lemma}
\noindent{P}roof:~~~~See the proof of \cite[Appendix C-A]{vy3}.
\qed

\begin{lemma} \label{lemma2}
There {exists} an auxiliary RV $U$ satisfying $Z-X-Y-U$ and
\begin{align}
\textstyle \sum_{\substack{ t=1 }}^nI(Z_{t};U_t) &= nI(Z;U), \label{ztut} \\
\textstyle \sum_{\substack{ t=1 }}^nI(Y_{t}(W);U_t) &= nI(Y;U), \label{ytut}\\
\textstyle \sum_{\substack{ t=1 }}^nI(X_{t}(W);U_t) &= nI(X;U). \label{xtut}
\end{align}
\end{lemma}
\noindent{Proof:~~~~See the {proof} in \cite[Appendix C-B]{vy3}}.
\qed

{We fix the auxiliary RV $U$ specified by Lemma \ref{lemma2}}. The next lemma plays a key role in the analysis of privacy-leakage, which will be seen later.
\begin{lemma} \label{lemma3}
It holds that
\begin{align}
I(Z^n;\bm{J},W) \ge n(R_I - (\delta + \delta_n)), \label{znjwri}
\end{align}
where $\delta_n = \frac{1}{n}(1+\delta\log M_I M_C M_G)$, and $\delta_n \downarrow 0$ as $n \rightarrow \infty$ and $\delta \downarrow 0$.
\end{lemma}
\noindent{P}roof:~~~~We {expand} the left-hand side in \eqref{znjwri} {as}
\begin{align}
I(Z^n;\bm{J},W) &\ge I(Z^n;W|\bm{J}) = H(W|\bm{J}) - H(W|Z^n,\bm{J})\nonumber \\
&\overset{\mathrm{(a)}}\ge H(W) - n\delta_n = \log M_I - n\delta_n\nonumber \\
&\overset{\mathrm{(b)}}\ge n(R_I - (\delta + \delta_n)),
\end{align}
where
\begin{enumerate}[label=(\alph*)]
	\setcounter{enumi}{0}
        \vspace{-3mm}
	\item follows because Fano's inequality is applied,
	\vspace{-3mm}
        \item is due to \eqref{4b}.
        \qed
\end{enumerate}

\medskip
\noindent{\em Analysis of Identification, Chosen- and Generated-Secret Key Rates}:~~~~
We begin with considering the join entropy of $E(W)=(W,S_G(W),S_C(W))$ as
\begin{align}
H&(E(W))= H(E(W)|Z^n,\bm{J}) + I(E(W);Z^n,\bm{J}) \nonumber \\
&\overset{\mathrm{(c)}}= {H(E(W)|\widehat{E(W)},Z^n,\bm{J}) + I(E(W);Z^n,\bm{J})} \nonumber \\
&\overset{\mathrm{(d)}}\le n\delta_n + I(E(W);\bm{J}) + I(E(W);Z^n|\bm{J}) \nonumber \\
&= n\delta_n + I(W;\bm{J}) + I(S_C(W),S_G(W);\bm{J}|W) \nonumber \\
&~~~+ H(Z^n|\bm{J})-H(Z^n|\bm{J},S_C(W),S_G(W),W) \nonumber \\
&\overset{\mathrm{(e)}}= n\delta_n + I(S_C(W),S_G(W);J(W)|W) \nonumber \\
&~~~+ H(Z^n|J(W))-H(Z^n|T(W)) \nonumber \\
&\overset{\mathrm{(f)}}\le n(\delta_n + \delta) + H(Z^n)-H(Z^n|T(W)) \nonumber \\
&= \sum_{\substack{ t=1 }}^n\Big\{H(Z_t) - H(Z_t|Z^{t-1},T(W))\Big\} + n(\delta+\delta_n) \nonumber \\
&= \sum_{\substack{ t=1 }}^nI(Z_t;U_t) + n(\delta+\delta_n)\nonumber \\
&\overset{\mathrm{(g)}}= n(I(Z;U) + \delta+\delta_n), \label{hriscsg}
\end{align}
where
\begin{enumerate}[label=(\alph*)]
	\setcounter{enumi}{2}
        \vspace{-3mm}
        \item {holds as $\widehat{E(W)}$ is a function of $(Z^n,\bm{J})$},
        \vspace{-3mm}
        \item {follows because conditioning reduces entropy}, and Fano's inequality is applied,
        \vspace{-3mm}
        \item holds because $W$ is independent of other RVs and only $J(W)$ is possibly dependent on $Z^n$, $S_C(W)$,
        and $S_G(W)$,
        \vspace{-6mm}
        \item follows because (\ref{4ch}) is used and conditioning reduces entropy,
        \vspace{-3mm}
        \item holds due to \eqref{ztut} in Lemma \ref{lemma2}.
\end{enumerate}
In the opposite direction, we can also derive the following relation:
\begin{align}
H(E(W)) &= H(W,S_C(W),S_G(W)) \nonumber \\
&= H(W) + H(S_C(W),S_G(W)|W) \nonumber \\
&\overset{\mathrm{(h)}}= H(W) + H(S_C(W)) + H(S_G(W)|W) \nonumber\\
&~~~-I(S_C(W);S_G(W)|W) \nonumber \\
&\overset{\mathrm{(i)}}= \log M_I + \log M_C
+ H(S_G(W)|W) \nonumber \\
&~~~-I(S_C(W);S_G(W)|W) \nonumber \\
&\overset{\mathrm{(j)}}\ge n(R_I + R_C + R_G - \Gamma - 3\delta), \label{hrircrg}
\end{align}
where
\begin{enumerate}[label=(\alph*)]
	\setcounter{enumi}{7}
        \vspace{-3mm}
        \item holds because $W$ and $S_C(W)$ are independent of each other,
        \vspace{-3mm}
        \item follows since $W$ and $S_C(W)$ {are} uniformly distributed on $\mathcal{I}$ and $\mathcal{S}_C$, respectively,
        \vspace{-3mm}
        \item is due to (\ref{4b}), \eqref{4c}, (\ref{4ce}), and (\ref{4cg}).
\end{enumerate}
From (\ref{hriscsg}) and (\ref{hrircrg}), we obtain
\begin{align}
R_I + R_C + R_G &\le I(Z;U) + \Gamma + 4\delta + \delta_n. \label{rcrglast}
\end{align}

Using \eqref{hriscsg}, it is straightforward that
\begin{align}
    H(W,S_C(W)) &\le H(W,S_C(W),S_G(W)) \nonumber\\
    &\le n(I(Z;U) + \delta+\delta_n),
\end{align}
and
\begin{align}
    H(W,S_G(W)) \le n(I(Z;U) + \delta+\delta_n).
\end{align}

By a similar manner of \eqref{hrircrg}, it can be shown that
$
    H(W,S_C(W))\ge n(R_I + R_C -2\delta)
$
and
$
    H(W,S_G(W))\ge n(R_I + R_G -2\delta)
$, and therefore,
\begin{align}
    R_I + R_C &\le I(Z;U) + 3\delta+\delta_n, \label{rircizu} \\
    R_I + R_G &\le I(Z;U) + 3\delta+\delta_n. \label{rirgizu}
\end{align}
From \eqref{rircizu} and \eqref{rirgizu}, one can easily see that
\begin{align}
    R_I + R_C + R_G &\le I(Z;U) + R_G + 3\delta+\delta_n, \label{rircrgizurg} \\
    R_I + R_G + R_C &\le I(Z;U) + R_C + 3\delta+\delta_n. \label{rirgrcizurc}
\end{align}
{Finally}, we have that
\begin{align}
    \hspace{-3mm}R_I + R_C + R_G \le I(Z;U) + \min\{\Gamma,R_G,R_C\} + 4\delta+\delta_n, \label{rircrgizu}
\end{align}
where Equation \eqref{rircrgizu} is due to comparing the values on the right-hand sides of
\eqref{rcrglast}, \eqref{rircrgizurg}, and \eqref{rirgrcizurc}, and the smallest one is {a}
valid bound for all these equations.

\medskip
\noindent{{\em {Analysis} of Storage Rate}}:~~~~We have that
\begin{align}
&n(R_J + \delta) \ge \log M_J \ge \max_{w \in \cal{I}}H(J(w)) \nonumber \\
&\ge \frac{1}{M_I}\sum_{\substack{ w=1 }}^{M_I}H(J(W)|W=w) = H(J(W)|W) \nonumber \\
&=I(Y^n_W,S_C(W);J(W)|W) \nonumber \\
&\overset{\mathrm{(k)}}=H(Y^n_W) + H(S_C(W))  \nonumber \\
&~~~~- H(Y^n_W,S_C(W),S_G(W)|J(W),W) \nonumber \\
&\overset{\mathrm{(l)}}=H(Y^n_W) - H(Y^n_W|T(W)) + \log M_C  \nonumber \\
&~~~~- H(S_C(W),S_G(W)|J(W),W) \nonumber \\
&\ge \sum_{\substack{ t=1 }}^n \Big\{H(Y_t(W)) - H(Y_t(W)|Y^{t-1}(W),T(W))\Big\}\nonumber \\
&~~~ + \log M_C - H(S_C(W),S_G(W)|W)\nonumber \\
&\overset{\mathrm{(m)}}= \sum_{\substack{ t=1 }}^n \Big\{H(Y_t(W)) - H(Y_t(W)|Z^{t-1},Y^{t-1}(W),T(W))\Big\}  \nonumber \\
&~~~~+ n(R_C-\delta) - H(S_C(W),S_G(W)|W) \nonumber \\
&\overset{\mathrm{(n)}}\ge \sum_{\substack{ t=1 }}^n \Big\{H(Y_t(W)) - H(Y_t(W)|Z^{t-1},T(W))\Big\}  \nonumber \\
&~~~~+ n(R_C-\delta) - n(I(Z;U)-R_I+2\delta + \delta_n) \nonumber \\
&= \sum_{\substack{ t=1 }}^nI(Y_t(W);U_t) - n(I(Z;U) - R_I - R_C +3\delta + \delta_n)\nonumber \\
&\overset{\mathrm{(o)}}=n(I(Y;U) - I(Z;U) + R_I + R_C - 3\delta - \delta_n), \label{nrjdelta}
\end{align}
where
\begin{enumerate}[label=(\alph*)]
	\setcounter{enumi}{10}
        \vspace{-3mm}
        \item holds because {$Y^n_W$ is independent of $S_C(W)$}, and $S_G(W)$ is a function of $(Y^n_W,S_C(W))$,
        \vspace{-3mm}
        \item holds since $S_C(W)$ is uniformly distributed on $\mathcal{S}_C$ and $W$ is independent of other RVs,
        \vspace{-3mm}
        \item follows due to (\ref{4c}) in Definition \ref{def11} and \eqref{zytw} in Lemma \ref{lemma1},
        \vspace{-7mm}
        \item follows as conditioning reduces entropy and from (\ref{hriscsg}), we have
        {$H(S_C(W),S_G(W)|W) \le n(I(Z;U) - R_I + 2\delta + \delta_n)$},
        \vspace{-3mm}
        \item holds due to \eqref{ytut} in {Lemma \ref{lemma2}}.
\end{enumerate}

\medskip
\noindent{\em {Analysis of Privacy-Leakage Rate}}:~~~~{From} \eqref{4cf},
{
\begin{align}
\hspace{-5mm} &n(R_L + \delta) \ge I(X^n_W;J(W)|W) \nonumber \\
&= I(X^n_W;J(W),S_C(W),S_G(W),Z^n|W) \nonumber \\
&~~~- I(X^n_W;S_C(W),S_G(W),Z^n|J(W),W) \nonumber \\
&= I(X^n_W;J(W),S_C(W),S_G(W)|W) \nonumber \\
&~~~+I(X^n_W;Z^n|J(W),S_C(W),S_G(W),W) \nonumber \\
&~~~- H(S_C(W),S_G(W),Z^n|J(W),W) \nonumber \\
&~~~+ H(S_C(W),S_G(W),Z^n|J(W),W,X^n_W) \nonumber \\
& \overset{\mathrm{(p)}}=  I(X^n_W;T(W))+H(Z^n|T(W))-H(Z^n|T(W),X^n_W) \nonumber \\
&~~~- H(Z^n|J(W),W)  \nonumber \\
&~~~- H(S_C(W),S_G(W)|J(W),W,Z^n) \nonumber \\
&~~~+ H(S_C(W),S_G(W)|J(W),W,X^n_W) \nonumber \\
&~~~+ H(Z^n|T(W),X^n_W) \nonumber \\
&\overset{\mathrm{(q)}}\ge I(X^n_W;T(W))-(H(Z^n)-H(Z^n|T(W))) \nonumber \\
&~~~+ H(Z^n) - H(Z^n|J(W),W)-n\delta_n \nonumber \\
&\overset{\mathrm{(r)}}= \sum_{\substack{ t=1 }}^n\{I(X_t(W);X^{t-1}(W),T(W))-I(Z_t;Z^{t-1},T(W))\} \nonumber \\
&~~~+ I(Z^n;J(W),W)-n\delta_n \nonumber \\
&\overset{\mathrm{(s)}}= \sum_{\substack{ t=1 }}^n\{I(X_t(W);Z^{t-1},X^{t-1}(W),T(W))-I(Z_t;U_t)\} \nonumber \\
&~~~+ I(Z^n;J(W),W)-n\delta_n \nonumber \\
&\ge \sum_{\substack{ t=1 }}^n\{I(X_t(W);Z^{t-1} ,T(W))-I(Z_t;U_t)\} \nonumber \\
&~~~+ I(Z^n;J(W),W)-n\delta_n \nonumber \\
&\overset{\mathrm{(t)}}\ge \sum_{\substack{ t=1 }}^n\Big\{I(X_t(W);U_t) - I(Z_t;U_t) \Big\} \nonumber \\
&~~~+n(R_I - (\delta+\delta_n))-n\delta_n\nonumber \\
&\overset{\mathrm{(u)}}= n(I(X;U) - I(Z;U) + R_I -\delta - 2\delta_n), \label{rllast}
\end{align}
where
\begin{enumerate}[label=(\alph*)]
	\setcounter{enumi}{15}
        \vspace{-3mm}
        \item follows because $W$ is independent of $X^n_W$,
        \vspace{-3mm}
        \item follows because the third term and the last term cancel out each other, Fano's inequality is applied for the fifth term, and the sixth term is eliminated, 
        \vspace{-3mm}
        \item follows since each symbol of the sequences $(X^n_W,Z^n)$ is i.i.d.\ generated, and thus $\sum_{t=1}^nI(X_t(W);X^{t-1}(W)) = \sum_{t=1}^nI(Z_t;Z^{t-1})=0$,
        \vspace{-2mm}
        \item is due to \eqref{zxtw},
        \vspace{-3mm}
        \item follows from Lemma \ref{lemma3},
        \vspace{-3mm}
        \item holds due to \eqref{ztut} and \eqref{xtut} in Lemma \ref{lemma2}.
\end{enumerate}
}

For the cardinality bound $|\mathcal{U}| \le |\mathcal{Y}| + 2$,
we can derive the condition by using the support lemma {\cite[Appendix C]{GK}}.
Finally, by letting $n \rightarrow \infty$ and $\delta \downarrow 0$, we complete the proof of the converse part.
\qed

\subsection*{B. {Direct Part (Achievability)}}
{In the proof, we show only the case where $\Gamma \le {\min\{R_C,R_G\}}$. The other cases, namely,
(I) $R_C \le {\min\{\Gamma,R_G\}}$ and (I\hspace{-.08em}I) $R_G \le {\min\{\Gamma,R_C\}}$,
the constraint $R_I + R_C + R_G \leq I(Z;U) + \min\{\Gamma,R_C,R_G\}$ in Theorem \ref{th1} reduces to
$R_I + R_G \leq I(Z;U)$ for (I) and $R_I + R_C \leq I(Z;U)$ for (II), respectively, which already
emerge in Theorem \ref{th1}.
The proof of these cases follows similarly to that of $\Gamma \le {\min\{R_C,R_G\}}$ with minor adjustments.
In this part, we omit the detailed proof
although how to prove cases (I) and (I\hspace{-.08em}I) will be mentioned later.}


\medskip
\noindent{\bf Parameter Settings}:~~~~
First, fix $P_{U|Y}$. Let $\delta$ be a small enough positive value and fix $n$. We set
\begin{align}
    \hspace{-3mm}R_I &> 0, R_C > 0~(R_I + R_C < I(Z;U)), \\
    \hspace{-3mm}R_G &= I(Z;U) + \Gamma - (R_I + R_C) - \delta, \\
    \hspace{-3mm}R_M &= I(Y;U) - I(Z;U) + R_I + 2\delta, \\
    \hspace{-3mm}R_J &= I(Y;U) - I(Z;U) + R_I + R_C + 2\delta, \\
    \hspace{-3mm}R_L &= I(X;U) - I(Z;U) + R_I + 2\delta,
\end{align}
where $R_M$ denotes the rate of {a} dummy message shared between the encoder and decoder. We set {$\mathcal{I}=[1:2^{nR_I}]$}, $\mathcal{S}_C = [1:2^{nR_C}]$, $\mathcal{S}_G = [1:2^{nR_G}]$, and $\mathcal{J} = [1:2^{nR_J}]$.
We {also} define four new sets $\mathcal{S}_{\Gamma} = [1:2^{n\Gamma}]$,
$\mathcal{S}_{C\backslash\Gamma} = [1:2^{n(R_C-\Gamma)}]$,
$\mathcal{S}_{G\backslash\Gamma} = [1:2^{n(R_G-\Gamma)}]$, and  $\mathcal{M} = [1:2^{nR_M}]$,
representing the sets of shared bits,
unshared bits in chosen- and generated-secret keys, and dummy message, respectively.
Without loss of generality, there {exist} one-to-one mapping {tables} between
$l \in \mathcal{S}_C$ and a pair $(m,n) \in \mathcal{S}_{\Gamma} \times \mathcal{S}_{{C\backslash\Gamma}}$,
and between $p \in \mathcal{S}_G$ and $(q,r) \in \mathcal{S}_{\Gamma} \times \mathcal{S}_{{G\backslash\Gamma}}$

\medskip
\noindent{\bf Codebook Generation}:~~~~
Generate $2^{n(I(Y;U)+\delta)}$ sequences of $u^n(s_1,s_2,m)$, which are i.i.d.\ from $P_{U}$,
where $s_1 \in \mathcal{S}_C$, $s_2 \in \mathcal{S}_{{G\backslash\Gamma}}$, and $m \in \mathcal{M}$.

\begin{figure}[!t]
\vspace{-4mm}
 \centering
  \includegraphics[width =80mm]{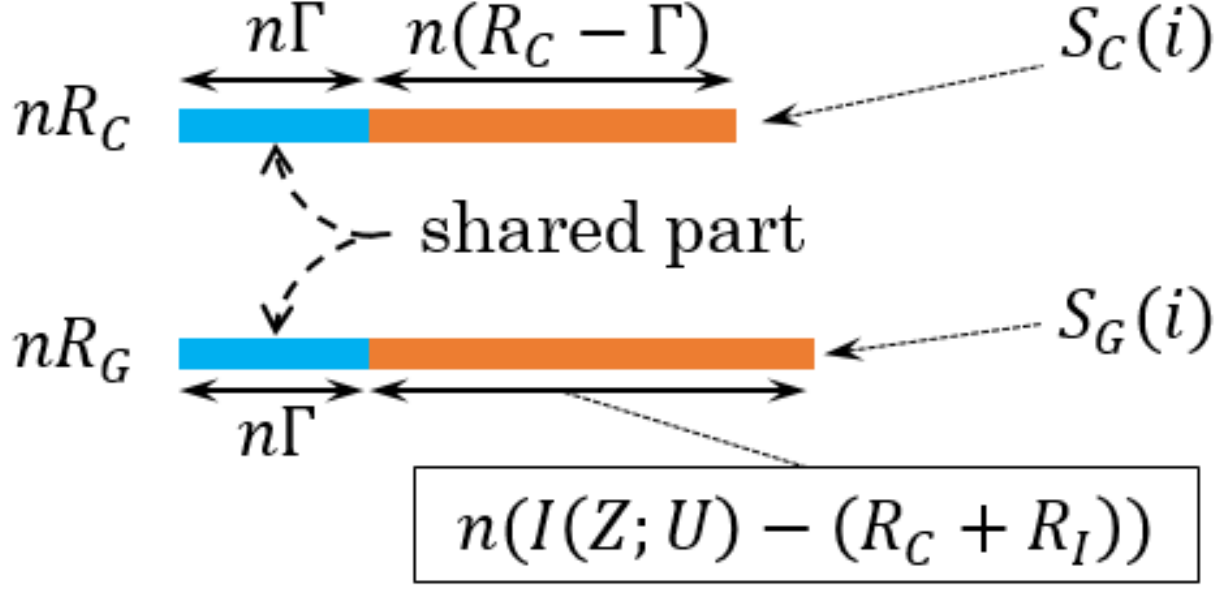}
   \vspace{-2mm}
 \caption{Shared bits: The blue parts {represent} information bits shared between chosen- and generated-secret keys.}
 \label{fig2222}
 \vspace{-4mm}
\end{figure}

\medskip
\noindent{\bf Encoding (Enrollment)}:~~~~
Note that
there is a one-to-one mapping
between $s_C(i)$ and a pair $(s_{C1}(i),s_{C2}(i))$, where $s_{C1}(i) \in \mathcal{S}_{\Gamma}$, and $s_{C2}(i) \in \mathcal{S}_{{C\backslash\Gamma}}$.
The first $n\Gamma$ information bits of $s_C(i)$, which {are} $s_{C1}(i)$, are shared with the generated-secret key as depicted in Fig.\ \ref{fig2222}.

Observing the measurement $y^n_i$ and {the chosen-secret key} $s_C(i)$,
the encoder finds {an} index {tuple} $(s_1,s_2,m)$ such that $(y^n_i,u^n({s_1},s_2,m)) \in \mathcal{T}_{\epsilon}^{(n)}(YU)$.
If there exists multiple tuples satisfying the joint typicality above, it picks one of them at random. {If there
exists no such tuple, error is declared}.
Let $(s_1(i),s_2(i),m(i))$ denote the tuple chosen for given $y^n_i$. Then, the encoder generates\footnote{To prove cases (I) and (I\hspace{-.1em}I),
it suffices to set $R_C = R_G = I(Z;U)-R_I-\delta$ and $|\mathcal{S}_{C\backslash\Gamma}| = {|\mathcal{S}_{G\backslash\Gamma}|} =1$
(singleton). More specifically, the generated-secret key is simply composed by sharing all bits of the chosen-secret key,
i.e., $s_G(i) = s_C(i)$, and the one-to-one mapping tables become redundant. The helper {data} and the generated-secret key are in simpler forms, that is,
$j(i) = (m(i),s_C(i)\oplus s_1(i)))$ and $s_G(i) = s_{C}(i)$.}
{a helper {data} and a secret key} as follows:
\begin{align}
j(i) &= (m(i),s_C(i)\oplus s_1(i))), \\
s_G(i) &= (s_{C1}(i),s_2(i)), \label{s2sc}
\end{align}
where $\oplus$ denotes the addition modulo $M_C$.
$j(i)$ and $s_G(i)$ are saved at location $i$ in the helper-{data} {DB} and {the other} secret-key {DB}, respectively.

\medskip
\noindent{\bf Decoding (Identification)}:~~~~
Seeing $z^n$, the decoder {looks for} the tuple $(s_1,s_2,m(i))$ such that $(z^n,u^n(s_1,s_2,m(i))) \in \mathcal{T}_{\epsilon}^{(n)}(ZU)$
for {some} $i \in \mathcal{I}$ and some $s_1 \in \mathcal{S}_C, s_2 \in \mathcal{S}_{{G\backslash\Gamma}}$. If $(i,s_1,s_2)$ is unique,
the decoder sets $(\widehat{s_1(w)},\widehat{s_2(w)},\widehat{m(w)})=(s_1,s_2,m(i))$. Otherwise, it declares error.
Assume that $(i,{s_1},s_2)$ {is} uniquely found. Then, the decoder outputs $\widehat{w}=i$ and the chosen-secret key as
\begin{align}
\widehat{s_C(w)} = {s_C(\widehat{w})} \oplus s_1(\widehat{w}) \ominus \widehat{s_1(w)}{,}
\end{align}
where {$s_C(\widehat{w}) \oplus s_1(\widehat{w})$} is the latter half of the helper {data} $j(\widehat{w})$ and $\ominus$ denotes the subtraction modulo $M_C$. After that, the decoder determines the corresponding pair $(\widehat{s_{C1}(w)},\widehat{s_{C2}(w)})$ from the one-to-one mapping {table}, and {uses} $\widehat{s_{C1}(w)}$ to estimate the generated-secret key as
$
\widehat{s_G(w)} = (\widehat{s_{C1}(w)},\widehat{s_2(w)}).
$

\medskip
\noindent{\bf Evaluations of Performance}:~~~~
We shall check that all the conditions \eqref{4a}--\eqref{4h} in Definition \ref{def11} satisfy {with a}
random codebook $\mathcal{C}_n = \{U^n(s_1,s_2,m): s_1 \in \mathcal{S}_C, s_2 \in \mathcal{S}_{{G\backslash\Gamma}}, m \in \mathcal{M}\}$.
We first introduce some useful lemmas, which are often used in the evaluations of these conditions, and
then dive into the core part of the discussion. Let $(S_1(i),S_2(i),M(i))$ denote the corresponding index tuple of
individual $i$ chosen by the encoder for given $Y^n_i$. For simplicity, $(S_1(i),S_2(i),M(i))$ and
the sequence $U^n(S_1(i),S_2(i),M(i))$ {are} represented by $V(i)$ and $U^n(V(i))$, respectively. {Also, define {$\epsilon_n = \frac{1}{n} + \delta\log|\mathcal{Y}|$}, and this quantity goes to zero as the block length $n$ tends to infinity and $\delta$ approaches zero}.

\begin{lemma} \label{keylem2}

It holds that
\begin{align}
H(Y^n_i|V(i),\mathcal{C}_n) &\leq n(H(Y|U) + {\epsilon_n}), \label{yjs1s2} \\
H(Y^n_i|X^n_i,V(i),\mathcal{C}_n) &\leq n(H(Y|X,U) + {\epsilon_n}). \label{yxjs1s2}
\end{align}
\end{lemma}
\noindent{Proof}:~~~~Since $V(i)$ determines {$U^n(V(i))$} directly,
the above lemma can be viewed as {\cite[Lemma 4]{kitti}}. For the detailed proof,
see \cite[Appendix C]{kitti} or \cite[Appendix B]{vy3} as a reference.
\qed
\begin{lemma} \label{keylem3}
We have that
\begin{align}
&\hspace{-3mm}H(S_1(i)|\mathcal{C}_n) \ge n(R_C - \delta - {\epsilon_n}), \label{keylem3-1} \\
&\hspace{-3mm}H(S_2(i)|\mathcal{C}_n) \ge n(R_G - \Gamma - \delta - {\epsilon_n}), \label{keylem3-2} \\
&\hspace{-3mm}H(S_1(i),S_2(i)|\mathcal{C}_n) \ge n(I(Z;U)-R_I - 2\delta - {\epsilon_n}), \label{keylem3-3} \\
&\hspace{-3mm}I(S_1(i),S_2(i);M(i)|\mathcal{C}_n) \le n(\delta +{\epsilon_n}), \label{keylem3-5} \\
&\hspace{-3mm}I(S_1(i);S_2(i),M(i)|\mathcal{C}_n) \le n(\delta +{\epsilon_n}). \label{keylem3-4}
\end{align}
\end{lemma}
\noindent{Proof}:~~~~The proof is provided in Appendix B.
\qed

\medskip
\noindent{\em Analysis of Error Probability}:~~~~
For $W=i$, {an} error event possibly happens at the encoder is
\begin{enumerate}[label={}]
	\item [$\mathcal{E}_1$]:~\{$(Y^n,U^n(s_1,s_2,m)) \notin \mathcal{T}_{\epsilon}^{n}(YU)$  for all $s_1 \in \mathcal{S}_C$, $s_2 \in$ \\
        ~~~~~$\mathcal{S}_{{G\backslash\Gamma}}$, $m \in \mathcal{M}$\},
\end{enumerate}
and {those} at the decoder are:
\begin{enumerate}[label={}]
	\item [${\mathcal{E}_2}$]:~\{$(Z^n,U^n(V(i))) \notin \mathcal{T}_{\epsilon}^{n}(ZU)$\},
        \vspace{-3mm}
	\item [${\mathcal{E}_3}$]:~\{$(Z^n,U^n(S_1(i),s'_2,M(i)) \in \mathcal{T}_{\epsilon}^{n}(ZU)$ for $\exists s'_2 \neq S_2(i)$,\\
        ~~~~~$s'_2 \in \mathcal{S}_{{G\backslash\Gamma}}$\},
	\vspace{-3mm}
        \item [${\mathcal{E}_4}$]:~\{$(Z^n,U^n(s'_1,S_2(i),M(i)) \in \mathcal{T}_{\epsilon}^{n}(ZU)$ for $\exists s'_1 \neq S_1(i)$, \\
        ~~~~~$s'_1 \in \mathcal{S}_C$\},
    \vspace{-3mm}
    \item[${\mathcal{E}_5}$]:~\{$(Z^n,U^n(s'_1,s'_2,M(i)) \in \mathcal{T}_{\epsilon}^{n}(ZU)$ for $\exists s'_1 \neq S_1(i)$, \\
    ~~~~~$s'_1 \in \mathcal{S}_C$, and $\exists s'_2 \neq S_2(i)$, $s'_2 \in \mathcal{S}_{{G\backslash\Gamma}}$\},
    \vspace{-3mm}
    \item [${\mathcal{E}_6}$]:~\{$(Z^n,U^n(s_1,s_2,M(i')) \in \mathcal{T}_{\epsilon}^{n}(ZU)$ for $\exists i' \neq i$, $i' \in$ \\
    ~~~~~$\mathcal{I}$,~$s_1 \in \mathcal{S}_C$, and $s_2 \in \mathcal{S}_{{G\backslash\Gamma}}$\}.
\end{enumerate}
{Note that $E(W) = (W,S_G(W),S_C(W))$, and then we can further {bound} the entire error probability as}
\vspace{-4mm}
\begin{align}
\Pr\Big\{&\widehat{E(W)} \neq E(W)|W=i\Big\} = \Pr\Big\{\bigcup_{i=1}^6\mathcal{E}_i\Big\} \nonumber \\
&\leq \Pr\left\{\mathcal{E}_1\right\} +
\Pr\left\{{\mathcal{E}_2|\mathcal{E}^c_1}\right\} \nonumber \\
&~~~+\Pr\left\{\mathcal{E}_3\right\}+\Pr\left\{\mathcal{E}_4\right\}+\Pr\left\{\mathcal{E}_5\right\}
+\Pr\left\{\mathcal{E}_6\right\}. \label{e1toe6}
\end{align}

By using the covering lemma \cite[Lemma 3.3]{GK}, $\Pr\{\mathcal{E}_1\}$ can be made smaller than $\delta$ since $R_C+R_G-\Gamma+R_M > I(Y;U)$. The $\Pr\left\{{\mathcal{E}_2|\mathcal{E}^c_1}\right\}$ can also be
made small enough by the Markov lemma \cite[Lemma 15.8.1]{cover}. {The probabilities} $\Pr\left\{\mathcal E_i\right\}$, $i=3,\cdots,6$ vanish as well by applying the packing lemma \cite[Lemma 3.1]{GK}
since we have $R_I \ge 0, R_C \ge 0, R_G \ge 0 $, and $R_I+R_C+R_G-\Gamma < I(Z;U)$. Overall, the error probability {averaged over the random codebook} can be bounded by
\begin{align}
\Pr\big\{\widehat{E(W)} \neq E(W)|W=i\big\} \leq 6\delta \label{errpro}
\end{align}
for large enough $n$.

\medskip
\noindent{\em Analyses of Identification and Chosen-{Secret Key} Rates}:~~~~
From the parameter settings, \eqref{4b} and \eqref{4c} are trivial.

\medskip
\noindent{\em Analysis of Generated-{Secret Key} Rate}:~~~~{For} the left-hand side of (\ref{4e}), we have that
\begin{align}
H(S_G(i)|\mathcal{C}_n) &= H(S_{C1}(i),S_2(i)|\mathcal{C}_n) \nonumber \\
&\overset{\mathrm{(a)}}= H(S_{C1}(i)|\mathcal{C}_n) + H(S_2(i)|\mathcal{C}_n) \nonumber\\
&\overset{\mathrm{(b)}}\ge n(R_G -\delta - {\epsilon_n}), \label{rgdelta}
\end{align}
where 
\begin{enumerate}[label=(\alph*)]
	\setcounter{enumi}{0}
        \vspace{-3mm}
	\item holds as $S_{C1}(i)$ is independent of $S_2(i)$,
        \vspace{-3mm}
	\item follows because $S_{C1}(i)$ is uniformly distributed on $\mathcal{S}_{\Gamma}$ and \eqref{keylem3-2} is applied.
\end{enumerate}

\medskip
\noindent{\em Analysis of Storage Rate}:~~~~
The total required storage rate is
\begin{align}
&\frac{1}{n}\log M_J \le R_M + R_C \nonumber \\
&= I(Y;U) - I(Z;U) + R_I + 2\delta + R_C \le R_J + \delta. \label{storage}
\end{align}

\medskip
\noindent{\em Analysis of Privacy-Leakage Rate}:~~~~By invoking the same arguments around \cite[Appendix B]{YYO2021a}, we obtain that
\begin{align}
I(X^n_i;J(i)|\mathcal{C}_n) = I(X^n_i;M(i)|\mathcal{C}_n). \label{xnmcn}
\end{align}
From (\ref{xnmcn}), {since we denote $V(i)=(S_1(i),S_2(i),M(i))$}, we have
\begin{align}
\hspace{-4cm}I&(X^n_i;M(i)|\mathcal{C}_n) \nonumber \\
&= I(X^n_i;V(i)|\mathcal{C}_n) - I(X^n_i;S_1(i),S_2(i)|M(i),\mathcal{C}_n) \nonumber \\
&= H(V(i)|\mathcal{C}_n) - H(V(i)|X^n_i,\mathcal{C}_n)  \nonumber \\
&~~~- H(S_1(i),S_2(i)|M(i),\mathcal{C}_n) \nonumber \\
&~~~+ H(S_1(i),S_2(i)|M(i),X^n_i,\mathcal{C}_n) \nonumber \\
&\overset{\mathrm{(c)}}\le H(V(i)|\mathcal{C}_n) - \{H(Y^n_i,V(i)|X^n_i,\mathcal{C}_n) \nonumber\\
&~~~- H(Y^n_i|X^n_i,V(i),\mathcal{C}_n)\} - \{H(S_1(i),S_2(i)|\mathcal{C}_n)\nonumber \\
&~~~-I(S_1(i),S_2(i);M(i)|\mathcal{C}_n)\} + {n{\delta_n}} \nonumber \\
&\overset{\mathrm{(d)}}\le n(I(Y;U) + \delta) - \{H(Y^n_i,V(i)|X^n_i,\mathcal{C}_n) \nonumber\\
&~~~- H(Y^n_i|X^n_i,V(i),\mathcal{C}_n)\} - \{H(S_1(i),S_2(i)|\mathcal{C}_n)\nonumber \\
&~~~-I(S_1(i),S_2(i);M(i)|\mathcal{C}_n)\} + {n{\delta_n}} \nonumber \\
&\overset{\mathrm{(e)}}\le n(I(Y;U) + \delta) - \{H(Y^n_i,V(i)|X^n_i,\mathcal{C}_n) \nonumber\\
&~~~- H(Y^n_i|X^n_i,V(i),\mathcal{C}_n)\} - n(I(Z;U)-R_I - 2\delta - {\epsilon_n}) \nonumber \\
&~~~+ n(\delta + {\epsilon_n}) + n{\delta_n} \nonumber \\
&\overset{\mathrm{(f)}}\le nI(Y;U) - nH(Y|X) + H(Y^n_i|X^n_i,V(i),\mathcal{C}_n) \nonumber \\
&~~~-n(I(Z;U) - R_I - {4\delta} - 2{\epsilon_n} - {\delta_n}) \nonumber \\
&\overset{\mathrm{(g)}}\le nI(Y;U) - nH(Y|X) + n(H(Y|X,U) + {\epsilon_n}) \nonumber \\
&~~~-n(I(Z;U) - R_I - {4\delta} - 2{\epsilon_n} - {\delta_n}) \nonumber \\
&= n(I(Y;U) - I(Y;U|X)) - n(I(Z;U) \nonumber \\
&~~~- R_I - {4\delta} - 3{\epsilon_n} - {\delta_n}) \nonumber \\
&= n(H(U)-H(U|Y) - H(U|X) + H(U|Y,X))\nonumber \\
&~~~- n(I(Z;U) - R_I - {4\delta} - 3{\epsilon_n} - {\delta_n})\nonumber \\
&\overset{\mathrm{(h)}}= n(I(X;U)-I(Z;U) + R_I + {4\delta} + 3{\epsilon_n} +{\delta_n}), \label{xnmc}
\end{align}
where
\begin{enumerate}[label=(\alph*)]
	\setcounter{enumi}{2}
        \vspace{-3mm}
        \item follows since $(S_1(i),S_2(i))-(M(i),X^n_i)-Z^n$ holds for given $\mathcal{C}_n$, we have that
        $H(S_1(i),S_2(i)|M(i),X^n_i,\mathcal{C}_n) \le H(S_1(i),S_2(i)|M(i),Z^n,\mathcal{C}_n) \le n{\delta_n}$, where the last inequality is obtained by Fano's inequality {with $\delta_n = \frac{1}{n}(1+\delta\log M_I M_C M_G)$},
        \vspace{-3mm}
        \item {is due to $H(V(i)|\mathcal{C}_n) \le
        H(S_1(i)|\mathcal{C}_n) + H(S_2(i)|\mathcal{C}_n)+
        H(M(i)|\mathcal{C}_n) \le n(I(Y;U) + \delta)$},
        \vspace{-3mm}
        \item follows because (\ref{keylem3-3}) and (\ref{keylem3-5}) in Lemma \ref{keylem3} are applied,
        \vspace{-3mm}
        \item follows as {$X^n_i$ and $Y^n_i$ are independent of $\mathcal{C}_n$},
        \vspace{-3mm}
        \item is due to (\ref{yxjs1s2}) in Lemma \ref{keylem2},
        \vspace{-3mm}
        \item holds due to the Markov chain $X-Y-U$ (cf.\ (\ref{theorem1})), {that is,} $H(U|Y,X)=H(U|Y)$.
\end{enumerate}

\medskip
\noindent{\em Analysis of Secret-Key Leakage}:~~~~The information leakage between the two secret keys can be bounded as follows:
\begin{align}
I&(S_C(i);S_G(i)|\mathcal{C}_n) \nonumber \\
&= I(S_{C1}(i),S_{C2}(i);S_{C1}(i),S_2(i)|\mathcal{C}_n)\nonumber \\
&= H(S_{C1}(i),S_{C2}(i)|\mathcal{C}_n) \nonumber \\
&~~~- H(S_{C1}(i),S_{C2}(i)|S_{C1}(i),S_2(i),\mathcal{C}_n)\nonumber \\
&\overset{\mathrm{(i)}}= H(S_{C1}(i),S_{C2}(i)|\mathcal{C}_n)-H(S_{C2}(i)|\mathcal{C}_n)\nonumber \\
&= H(S_{C1}(i)|\mathcal{C}_n) + H(S_{C2}(i)|\mathcal{C}_n) -H(S_{C2}(i)|\mathcal{C}_n) \nonumber \\
&= n\Gamma, \label{scisgi}
\end{align}
where (i) follows as {$S_{C2}(i)$ is independent of $(S_{C1}(i),S_2(i))$.}

\medskip
\noindent{\em Analysis of Secrecy-Leakage}:~~~~{For (\ref{4h})}, it follows that
\begin{align}
I&(S_C(i),S_G(i);J(i)|\mathcal{C}_n) \nonumber \\
&\overset{\mathrm{(j)}}=I(S_C(i),S_2(i);M(i),S_C(i)\oplus S_1(i)|\mathcal{C}_n) \nonumber \\
&= H(M(i),S_C(i)\oplus S_1(i)|\mathcal{C}_n) \nonumber \\
&~~~ -H(M(i),S_C(i)\oplus S_1(i)|S_C(i),S_2(i),\mathcal{C}_n) \nonumber \\
&= H(M(i)|\mathcal{C}_n) + H(S_C(i)\oplus S_1(i)|M(i),\mathcal{C}_n) \nonumber \\
&~~~- H(M(i)|S_C(i),S_2(i),\mathcal{C}_n) \nonumber \\
&~~~- H(S_C(i)\oplus S_1(i)|M(i),S_C(i),S_2(i),\mathcal{C}_n) \nonumber \\
&\le H(M(i)|\mathcal{C}_n) + nR_C - H(M(i)|S_C(i),S_2(i),\mathcal{C}_n)  \nonumber \\
&~~~ - H(S_1(i)|M(i),S_C(i),S_2(i),\mathcal{C}_n) \nonumber \\
&\overset{\mathrm{(k)}}= H(M(i)|\mathcal{C}_n) + nR_C - H(M(i)|S_2(i),\mathcal{C}_n) \nonumber \\
&~~~ - H(S_1(i)|M(i),S_2(i),\mathcal{C}_n) \nonumber \\
&= nR_C- H(S_1(i)|\mathcal{C}_n) + I(S_2(i);M(i)|\mathcal{C}_n) \nonumber \\
&~~~+ I(S_1(i);S_2(i),M(i)|\mathcal{C}_n) \nonumber \\
&\overset{\mathrm{(l)}}\le 3n(\delta + {\epsilon_n}), \label{iscshj}
\end{align}
where
\begin{enumerate}[label=(\alph*)]
	\setcounter{enumi}{9}
        \vspace{-3mm}
	\item {is} due to the fact that $S_G(i) = ({S_{C1}(i)},S_2(i))$ and {$S_{C1}(i)$} is the {first} half of the chosen-secret key $S_C(i)$,
	\vspace{-3mm}
	\item holds since $S_C(i)$ is independent of other RVs,
	\vspace{-3mm}
    \item follows because (\ref{keylem3-1}), (\ref{keylem3-5}), and (\ref{keylem3-4}) in Lemma \ref{keylem3} are applied.
\end{enumerate}

By applying the selection lemma \cite[Lemma II]{BB} to all results shown above
(i.e., Eqs.\ (\ref{errpro}), (\ref{rgdelta}), (\ref{storage}), and (\ref{xnmc})--(\ref{iscshj})),
there exists at least {one} good codebook satisfying all the conditions in Definition \ref{def11} for large enough $n$.
\qed

\section*{Appendix B: Proof of Lemma \ref{keylem3}}
We begin with checking (\ref{keylem3-1}). In view of $V(i) = (S_1(i),S_2(i),M(i))$, we have
\begin{align}
H&(S_1(i)|\mathcal{C}_n) \nonumber \\
&= H(V(i)|\mathcal{C}_n) - H(S_2(i),M(i)|S_1(i),\mathcal{C}_n) \nonumber \\
&= H(Y^n_i,V(i)|\mathcal{C}_n) - H(S_2(i),M(i)|S_1(i),\mathcal{C}_n)\nonumber \\
&~~~ - H(Y^n_i|V(i),\mathcal{C}_n) \nonumber \\
&\overset{\mathrm{(a)}}\ge H(Y^n_i) - H(S_2(i)|\mathcal{C}_n) -H(M(i)|\mathcal{C}_n) \nonumber \\
&~~~- H(Y^n_i|V(i),\mathcal{C}_n) \nonumber \\
&\overset{\mathrm{(b)}}\ge nH(Y) - n(I(Z;U)-R_I-R_C-\delta) \nonumber \\
&~~~-n(I(Y;U)-I(Z;U)+R_I+2\delta) \nonumber \\
&~~~- n(H(Y|U) + {\epsilon_n}) \nonumber \\
&~{=}~n(R_C - \delta - {\epsilon_n}), \label{rcdelta}
\end{align}
where
\begin{enumerate}[label=(\alph*)]
\setcounter{enumi}{0}
\vspace{-3mm}
\item follows because conditioning reduces entropy and $Y^n_i$ is independent of $\mathcal{C}_n$,
\vspace{-3mm}
\item follows because \eqref{yjs1s2} in Lemma \ref{keylem2} is applied.
\end{enumerate}
Similar to the arguments around (\ref{rcdelta}), it can be verified that
(\ref{keylem3-2}) and (\ref{keylem3-3}) hold, and therefore we omit {the details}.

For (\ref{keylem3-5}), it can be shown that
\begin{align}
I&(S_1(i),S_2(i);M(i)|\mathcal{C}_n) \nonumber \\
&= H(S_1(i),S_2(i)|\mathcal{C}_n) + H(M(i)|\mathcal{C}_n) - H(V(i)|\mathcal{C}_n) \nonumber \\
&= H(S_1(i),S_2(i)|\mathcal{C}_n) + H(M(i)|\mathcal{C}_n) \nonumber \\
&~~~- H(Y^n_i,V(i)|\mathcal{C}_n) + H(Y^n_i|V(i),\mathcal{C}_n) \nonumber \\
&\overset{\mathrm{(c)}}\le H(S_1(i)|\mathcal{C}_n) + H(S_2(i)|\mathcal{C}_n) + H(M(i)|\mathcal{C}_n)\nonumber \\
&~~~- H(Y^n_i) + H(Y^n_i|V(i),\mathcal{C}_n) \nonumber \\
&\overset{\mathrm{(d)}}\le nR_C + n(I(Z;U)-R_I - R_C -\delta) \nonumber \\
&~~~+ n(I(Y;U) - I(Z;U) + R_I + 2\delta) \nonumber \\
&~~~- nH(Y) + n(H(Y|U) + {\epsilon_n}) \nonumber \\
&=n(\delta +{\epsilon_n}), \label{s1s2m}
\end{align}
where
\begin{enumerate}[label=(\alph*)]
\setcounter{enumi}{2}
\vspace{-3mm}
\item follows because conditioning reduces entropy,
\vspace{-3mm}
\item follows because \eqref{yjs1s2} in Lemma \ref{keylem2} is applied.
\end{enumerate}
{Equation} \eqref{keylem3-4} can be {shown} in a similar manner.
\qed

\profile{Vamoua Yachongka}{received his Ph.D.\ degree in communication engineering and informatics from the University of Electro-Communications, Japan, in March 2021. He is currently a postdoc researcher at the same institution. His research interests include information theory and information-theoretic security.
}

\profile{Hideki Yagi}{received the B.E. degree, the M.E.\ degree, and the Ph.D. degree in industrial
and management systems engineering from Waseda University, Tokyo, Japan in 2001, 2003, and 2005, respectively.
He is currently an Associate Professor at the Department of Computer and Network Engineering,
University of Electro-Communications, Tokyo, Japan.
He was with Media Network Center, Waseda University as a Research Associate from 2005 to 2007,
and an Assistant Professor from 2007 to 2008.
In the winter of 2008 and from July, 2010 to January, 2011, he was a Visiting Fellow at Princeton
University. His research interests include information and coding theory and information theoretic
security. Dr.\ Yagi is a member of Research Institute of Signal Processing and IEEE.
}
\end{document}